\documentstyle[aps,prl,multicol,epsf]{revtex}
\begin{document}
\draft
\title{ Symmetrizing Evolutions
}
\author{Paolo Zanardi$^{1,2}$
\footnote{Electronic address: zanardi@isiosf.isi.it}
} 
\address{
$^{1}$ Istituto Nazionale per la Fisica della Materia (INFM) \\
$^{2}$ Institute for Scientific Interchange  Foundation, \\Villa Gualino,
Viale Settimio Severo 65, I-10133 Torino, Italy\\
}
\date{\today}
\maketitle
\begin{abstract}
We introduce quantum procedures
for making $\cal G$-invariant the dynamics of 
 an arbitrary quantum system $S,$ where $\cal G$ is a finite  group acting on the space state of $S.$
Several applications of this idea are discussed.
In particular  
when $S$ is a $N$-qubit quantum computer interacting with its environment
and $\cal G$ the symmetric group of qubit permutations,
the resulting effective dynamics 
admits  noiseless subspaces. Moreover it is shown
that the recently introduced iterated-pulses schemes for reducing decoherence
in quantum computers fit in this general framework.
The noise-inducing component of the Hamiltonian is filtered out by the symmetrization
procedure just due to its transformation properties.
\end{abstract}
\pacs{PACS numbers: 03.67.Lx, 03.65.Fd}
\begin{multicols}{2}
\narrowtext

The importance of the notion of symmetry in quantum theory cannot be overstimated \cite{CORN}. 
The associated state-space decomposition into dynamically invariant sectors
is a highly desirable property in that it can strongly simplify the analysis
of the  system evolution. Suppose that on 
the state-space $\cal H$ of a quantum system $S$ 
acts a  group $\cal G$ via  a representation $\rho.$
In general the Hamiltonian $H$ of $S$ is not $\cal G$-invariant i.e., $[H,\,\rho({\cal G})]\neq 0.$
The  goal of this letter is to present a  quantum procedure for generating
an effective dynamics on $\cal H$ ruled by an operator  $\tilde H$ that
is the   $\cal G$-invariant component of $H.$
It amounts to a sort of generalized Fourier transform is which one discards
all the non-zero (i.e., non-translation invariant) components.
We first discuss a procedure 
that involves frequently iterated measurements.
The key idea is very simple: introducing an auxiliary space and resorting
to the intrinsic parallelism of  quantum dynamics  one can simultaneously evolve
all the group-rotated copies of an initial state.
Then by repeated measurements one singles out the $\cal G$-invariant component of the dynamics. 
After discussing several applications to state preparation, decoherence avoiding/suppression
and constrained dynamics, we show that  symmetrization can be achieved by purely unitary means
and without additional space resources.
This formulation  will make apparent that the recently proposed schemes for decoherence
control \cite{LLO}, \cite{VITO} in quantum computers \cite{QC} are nothing but special cases
of this general group-theoretic idea.
For the sake of clarity in this letter
we will concentrate on physical examples mostly suggested by quantum computation.
A deeper analysis of the algebraic structures involved 
along with further applications will be presented elsewhere.

Let us begin by a simple example aimed to give a first  hint about the
possible use of  $\cal G$-symmetrization for noise suppression.
Let $S$ be a single two-level system ({\em qubit}) dissipatively coupled  with an environment $E$.
${\cal H}={\bf{C}}^2\otimes{\cal H}_E,$ and $H=H_0+H_1,$ where
\begin{equation}
%$
 H_0=\varepsilon\,\sigma_z\otimes\openone +{\openone}\otimes H_E,\,H_1=
\sigma^+\otimes E+ \sigma^-\otimes E^\dagger.
%$
\end{equation}
Here $H_E$ (${\cal H}_E$) is the environment Hamiltonian (state-space) and $E,\,E^\dagger$ 
operators associated to the creation/annihilation of elementary
excitations of $E.$
On the total  space acts the group $\{ g_0={\openone},\,g_1=\sigma^z\otimes{\openone}\}\cong {\cal Z}_2.$
The operators transform according the adjoint action: $X\mapsto g_\alpha^\dagger\,X\,g_\alpha,\,
(\alpha=0,\,1).$
It is immediate to check that whereas the first two terms in $H$ (the self-Hamiltonians)
are invariant under the action of $\sigma^z,$
the interaction part changes sign ($\sigma^z\,\sigma^\pm\,\sigma^z=-\sigma^\pm$).
Therefore by `` averaging over the group '' $H$ one finds 
$\tilde H =2^{-1}\sum_{\alpha} g_\alpha^\dagger\,H\,g_\alpha=H_0.$
This tell us that if one, in some way, were able to make the system evolving
according  $\tilde H$ the interaction with the environment would be washed out.  

{\em Invariant subspaces.}
Now we  set the general framework and recall the relevant
group/representation-theoretic notions \cite{CORN}.
The general situation can be abstractly defined in terms of the data $({\cal H},\,H,\,{\cal G},\,\rho)$
where i) ${\cal H}$ is a finite dimensional Hilbert space, ii)
$H$ an hermitian operator (Hamiltonian) over ${\cal H}$,
iii) $\cal G$ a finite group  of order $|{\cal G}|,$ iv)
$\rho\,\colon \,g\in{\cal G}\mapsto \rho_g=\exp({i\,h_g^\rho}),$
a unitary representation of $\cal G$ in $\cal H$ [ $\rho_{gh}=\rho_{g}\,\rho_h,\,\rho_{g^{-1}}=
\rho_g^\dagger$]. 
The representation $\rho$ is irreducible  ({\em irrep}) if it does not admit
non-trivial invariant subspaces in $\cal H.$ 
The space $\cal H$ splits according the  $\cal G$-irreps:
${\cal H}= \oplus_J n_J\,{\cal H}_J $ where $n_J$  is the multiplicity 
of invariant subspace ${\cal H}_J$ associated to the $J$-th irrep of 
$\cal G.$
For instance the abelian (additive) group ${\cal Z}_2=\{0,\,1\}$ has two ($1$-$d$) irreps
$\rho_J(\alpha)=e^{i\,J\,\pi\alpha},$ the identical ($J=0$) and the antisymmetric one ($J=1$).

 In this letter we shall mainly focus on  the sector 
corresponding to the identity irrep. 
This is the subspace spanned by the vectors in ${\cal H}$
invariant under the action  of $\cal G:$
\begin{equation}
{\cal H}^\rho_{inv}=:\{ |\psi\rangle\in{\cal H}\,\colon\,
\rho_g\,|\psi\rangle=|\psi\rangle,\,\forall g\in{\cal G}\}.
\end{equation}
It is  easy  to check that 
the operator 
\begin{equation}
\pi_\rho=: {|{\cal G}|}^{-1} \sum_{g\in{\cal G}} \rho_g
\end{equation}
is  the projector onto ${\cal H}^\rho_{inv}$ \cite{CORN}.
In the very same way of all projections, $\pi_\rho$ has a
clear geometrical meaning:
from the elementary property 
$ \|\pi_\rho\,|\psi\rangle-|\psi\rangle\|=
\min_{|\phi\rangle\in{\cal H}^\rho_{inv}}\| |\psi\rangle -|\phi\rangle\|,$ 
 it follows
that $\pi_\rho\,|\psi\rangle$ represents the optimal $\cal G$-invariant
approximation of $|\psi\rangle$ and $\|\pi_\rho\,|\psi\rangle\|$
is a measure of the degree of $\cal G$-invariance of the vector $|\psi\rangle.$

Since in the following it will play the role of  {\em ancilla},
we  consider the so-called {\em Group Algebra} ${\bf{C}}{\cal G}$ of $\cal G.$
 It is a $|{\cal G}|$-dimensional vector space generated
by an orthonormal basis $\{|g\rangle\}$
 that is in a one-to-one correspondence with the elements of $\cal G.$
The  following two elements also will have  a major role in this paper
\begin{equation}
|0\rangle=: |{\cal G}|^{-1/2}\sum_{g\in{\cal G}} |g\rangle,
\quad W_\rho=: \sum_{g\in{\cal G}} \rho_g\otimes \Pi_g,
\end{equation}
where $ \Pi_g=: |g\rangle\langle g|.$ 
It is immediate 
to check that $W_\rho=e^{i\,K_\rho}$ is an  unitary operator
over  $ {\cal H}\otimes {\bf{C}}{\cal G},$ with generator  given by \cite{K}
%\begin{equation}
$
K_\rho=\sum_{g}h^\rho_g\otimes \Pi_g \label{infini}.
$ 
%\end{equation}
The physical meaning of the entangling operator $W_\rho$
should be quite  clear: it performs, 
conditionally on the group element encoded in the ancillary factor,
 the associated unitary rotations in computational space $\cal H.$
Concerning $|0\rangle$
we observe that also the appearance of this vector is very natural
in that the uniform superposition structure makes it  the unique  $\cal G$-invariant
element of the group algebra.

With these two ingredients one can design a simple quantum algorithm
for extracting the ${\cal G}$-invariant component of $|\psi\rangle.$
Let  $|\psi\rangle$ an arbitrary element of $\cal H.$
Apply $W_\rho$  to the initial state
$|\Psi_0\rangle=: |\psi\rangle\otimes |0\rangle:$
\begin{equation}
W_\rho\,|\Psi_0\rangle =\frac{1}{\sqrt{|{\cal G}|}}\sum_{g\in{\cal G}}
\rho_g\,|\psi\rangle\otimes |g\rangle.
\end{equation}
By projecting over $|0\rangle$ i.e., applying ${\openone}\otimes \Pi_0.$ 
one finds
\begin{equation}
|\Psi_0\rangle \mapsto |{\cal G}|^{-1}\sum_{g\in{\cal G}} \rho_g\,|\psi\rangle
\otimes |0\rangle = (\pi_\rho\otimes{\openone})|\Psi_0\rangle.
\end{equation}
Discarding the ancillary factor  one gets the ${\cal G}$-invariant component of $|\psi\rangle$
with  probability of success given by
$\|({\openone}\otimes \Pi_0)\,W_\rho\,|\Psi_0\rangle\|^2 =\|\pi_\rho\,|\psi\rangle\|^2.$
The procedure is illustrated by the following commutative diagram
\\
\def\normalbaselines{\baselineskip20pt\lineskip3pt\lineskiplimit3pt}
\def\mapright#1{{\mathop{\longrightarrow}\limits^{#1}}}
\def\mapup#1{\Big\uparrow\rlap{$\vcenter{\hbox{$\scriptstyle#1$}}$}}
 \def\mapdown#1{\Big\downarrow\rlap{$\vcenter{\hbox{$\scriptstyle#1$}}$}}
\begin{eqnarray}
\matrix{{\cal H}\otimes{\bf{C}}{\cal G} & \mapright{W_\rho}
 & {\cal H}\otimes{\bf{C}}{\cal G} \cr
        \mapup{{\openone}\otimes |0\rangle} && \mapdown{ {\openone}\otimes\Pi_0} \cr
        {\cal H} & \mapright{ \pi_{\rho} } & {\cal H} &}
        \nonumber
\end{eqnarray}
{\em Example 0} Let ${\cal H}={\cal H}_c^{\otimes\,N}$ 
a $N$-partite quantum system, ${\cal G}={\cal S}_N$
the symmetric group 
and $\rho$ the natural action of permutations on a tensor product
[$\rho(\sigma)\,\otimes_{j=1}^N |j\rangle=\otimes_{j=1}^N |\sigma(j)\rangle$].
Then $\pi_\rho(|\psi\rangle)$ is the totally symmetric component of $|\psi\rangle.$
Here we have an exponentially large ancilla ($|{\cal S}_N|=N!$). Any
permutation can be realized by a sequence of transpositions 
$t_{ij}\,|\phi\rangle_i\otimes|\psi\rangle_j= |\psi\rangle_i\otimes|\phi\rangle_j.$
In the 
qubit case i.e., ${\cal H}_c={\bf{C}}^2$
the $\{t_{ij}\}_{i,j=1}^N$  can be implemented in ${\cal H}$ by switching on, for a suitable
time, the two-qubit Hamiltonians $H_{ij}={\bf{s} }_i\cdot {\bf{s}}_j\,
[{\bf{s}}_i=:(\sigma_i^x, \sigma_i^y, \sigma_i^z)] .$ 

The described procedure can be immediately extended to the general $J$-th irrep of $\cal G.$
The corresponding projectors 
are given by $\pi_\rho^J= d_J/|{\cal G}|\sum_g \chi^{J\,*}(g) \,\rho_g,$ \cite{CORN}
where  $\chi^J =:\mbox {tr}\,\rho^J_g$
 ($d_J$) is the character (dimension) of the  $J$-th irrep. 
Now one has to project over $|J\rangle=:|{\cal G}|^{-1} \sum_g \chi^{J}(g)\, |g\rangle,$
eventually obtaining $d_J^{-1}\, \pi_\rho^J\,|\psi\rangle\otimes |J\rangle.$
This result  is useful, for example, in providing  a preparation procedure for
the $sl(d)$-singlets introduced in ref. \cite{ZARA} for noiseless quantum encoding
against collective decoherence in quantum computers.

{\em Example 1} Let ${\cal H},\,\cal G$ and $\rho$  as in  {\em Example. 0},
with ${\cal H}_c={\bf{C}}^d$ and $N= m\, d\,(m\in{\bf{N}}).$
Then there exists a (unique) ${\cal S}_N$-irrep $\bar J$ associated with the rectangular
Young tabelaux with $d$ rows.
$\pi_\rho^{\bar J}$ is the projector over the {\em singlet} sector
of $N$-fold tensor power of the defining irrep of $sl(d)$ \cite{CORN}.

Next example  shows how a simple group-theoretic structure is associated to any linear subspace.

{\em Example 2} Let $P$ be a projector in $\cal H,$
${\cal G}
=\{0,\,1\}\cong{\cal Z}_2,$ and $\rho\colon\alpha\mapsto e^{i\,\pi\,\alpha\,P}\, (\alpha=0, 1).$
One finds $\pi_\rho=2^{-1}({\openone}+  e^{i\,\pi\,P})=1-P=: P^\perp.$
Now ${\cal H}_{inv}^\rho$ is the the null subspace of $P.$
Conversely given a (non-trivial) representation $\rho$ of ${\cal Z}_2$ the space $\cal H,$
splits in the two orthogonal subspaces associated to the ${\cal Z}_2$-irreps:
$\rho(1)$ is a {\em parity} operator.
In the simplest instance of this situation,  when ${\cal H}={\bf{C}}^2$ and $P^\perp =  |0\rangle\langle 0|,$
 one has $K_\rho=|1\rangle\langle 1|\otimes |1\rangle\langle 1|,$ in terms of the Pauli operator 
$\sigma^z$ (and neglecting a trivial shift) this reads 
$2\,K_\rho= \sigma^z\otimes{\openone}+ {\openone}\otimes \sigma^z +1/2\, \sigma^z\otimes  \sigma^z.$
This expression shows that the interactions required for generating the unitary $W_\rho$
can be  physically reasonable.

{\em Unitary evolutions.}
Given the representation $\rho$ 
one can transform operators via  the adjoint action:
%\begin{equation}
$
\tilde\rho_g\colon X\mapsto \rho_g^\dagger\,X\,\rho_g.
$
%\end{equation}
The  subspace of $\cal G$-invariant operators is  then defined in the obvious way.
Now we present a  procedure for $\cal G$-symmetrizing unitary evolutions.
This is the natural operator extension of the projection/preparation procedures discussed above
and similarly it involves the group algebra as ancillary space and repeated measurements.

Let  $|\psi\rangle$ be an arbitrary element of $\cal H.$

 I) Apply $W_\rho$ to the initial state 
$|\Psi_0\rangle=: |\psi\rangle\otimes |0\rangle:$
\begin{equation} 
W_\rho\,|\Psi_0\rangle =\frac{1}{\sqrt{|{\cal G}|}}\sum_{g\in{\cal G}}
\rho_g\,|\psi\rangle\otimes |g\rangle.
\end{equation}

II) Evolve infinitesimally by $H\otimes {\openone},$
 i.e., apply $\delta U\otimes{\openone}$ where
$\delta U\simeq  {\openone}  -i\,\delta t\, H\;(\delta t=: t/M)$

III) Apply $W_\rho^\dagger,$ 
\begin{equation}
\frac{1}{\sqrt{|{\cal G}|}}\sum_{g\in{\cal G}} \rho_g^\dagger\,\delta U\,\rho_g\,|\psi\rangle
\otimes |g\rangle.
\end{equation}

IV) Project on $|0\rangle$
\begin{equation}
 \frac{1}{|{\cal G}|}\sum_{g\in{\cal G}} \rho_g^\dagger\,\delta U\,\rho_g\,|\Psi_0\rangle
\stackrel {\delta t\mapsto 0}{\simeq} 
({\openone} -i\,\delta t  \tilde H)\,|\Psi_0\rangle.
\end{equation}
Here 
%\begin{equation}
$
 \tilde H =:  |{\cal G}|^{-1} \sum_{g\in{\cal G}} \rho_g^\dagger \,H \,\rho_g=
\pi_{\tilde\rho}(H)
$
%\end{equation}
is by construction $\cal G$-invariant. 

V) Iterate of  I)--IV) $M$-times with $M\mapsto \infty.$

Steps I--IV amount to the  operation 
$T(\rho_0)=: S\,\rho_0\,S^\dagger,\,
%\begin{equation}
S= ({\openone}\otimes \Pi_0)\, W_\rho^\dagger\,(\delta U\otimes {\openone})\,W_\rho,$
with  $\rho_0=|\Psi_0\rangle\langle\Psi_0|.$%\end{equation}
The overall success probability  is given by
\begin{eqnarray}
& &\mbox{tr}\, T^M(\rho_0)= \mbox{tr}\,( S^M\,\rho_0\,S^{\dagger\,M})=
\|S^M\,|\Psi_0\rangle\|^2\nonumber \\ & &\stackrel{M\mapsto\infty}{\simeq}
\|({\openone}-\delta t\,\tilde H)^M\,|\psi\rangle\|^2\simeq
 \| e^{-i\,t\,\tilde H}\,|\Psi_0\rangle\|^2=1
\end{eqnarray}
The global evolution is then 
$\rho_0\mapsto T^M(\rho_0)/\mbox{tr}\, T^M(\rho_0) \simeq  S^M\,\rho_0\,S^{\dagger\,M},$
but
\begin{eqnarray}
& &S^M\,|\Psi_0\rangle  = [({\openone} - \frac{i\,t}{M}\,  \tilde H)^M\otimes {\openone}]\,|\Psi_0\rangle
\stackrel{M\mapsto\infty}{\longrightarrow}
\nonumber \\
& &(e^{-i\,t\,\tilde H}\otimes {\openone})\,|\Psi_0\rangle=
e^{-i\,t\,\tilde H}\,|\psi\rangle\otimes |0\rangle
\end{eqnarray}
Summarizing the above procedure [in the limit $M\mapsto\infty$] induces, in the
computational factor,  an effective dynamics generated by the $\cal G$-invariant Hamiltonian
$\tilde H=\pi_{\tilde\rho} (H).$ 
As argued above, $\tilde H$ represents the optimal $\cal G$-invariant
 approximation of $H$; from this point of view one can say that $\tilde U_t =e^{-i\,t\,\tilde H}$
is the natural {\em unitary}  symmetrization of $U_t.$ 
One has to  exploit a sort of quantum Zeno effect \cite{ZEN} [repeatedly measuring $|0\rangle$]
in that in order to obtain an admissible  quantum dynamics  in the
computational factor,  evolution has to be symmetrized  any infinitesimally small
amount of time. 
For example the naively symmetrized evolution $\tilde U =|{\cal G}|^{-1} \sum_g
\rho_g\,U\,\rho_g^\dagger  =:\pi_{\tilde\rho}(U)$ is not allowed, in that
it is not unitary. The symmetrization has to be ``exponentiated''.
%\cite{Naive}
If in step IV) projection over $|0\rangle$
were replaced by projection over $|J\rangle$ followed by the application of the unitary extension
of $|0\rangle\langle J|,$ eventually one would obtain the effective Hamiltonian
$H^J=:\sum_g \chi^{J*}(g)\rho^\dagger_g\,H\,\rho_g,$ that
 transforms according the $J$-th irrep of $\cal G.$

{\em{Example 3}}
With data like in {\em Ex. 2} one finds $\tilde H = P^\perp\,H\,P^\perp+  P\,H\,P.$
This shows that constraining the dynamics, by measurements, to a subspace is
a very special case of the general procedure introduced.
Notice that the projection measurement are over a single qubit ancilla.

{\em{Example 4}}:
 ${\cal H}= {\cal H}_c^{\otimes\,N}\otimes {\cal H}_{E},\, {\cal G}= {\cal S}_N,$
$\rho$ is the natural action over the first factor (like in {\em Ex. 0})
times the identity in ${\cal H}_E.$
The   dynamics thus obtained is {\em replica symmetric.}
This case is, in principle, relevant for quantum computation.
Indeed let us suppose that the computational factor is   a quantum register
made of  $N$ cells with state-space ${\cal H}_c$,
and ${\cal H}_{E}$ the state-space of environment.
Then the resulting (permutation invariant) effective  dynamics
admits decoherence-free subspaces suitable for noiseless quantum encoding \cite{ZARA},\cite{LID}.  
The minimal implementation of this example would require a setup
consisting of two qubits (interacting with an environment) and a third ancillary
qubit (coding for the symmetric group ${\cal S}_2$).
By performing the above procedure the singlet $2^{-1/2}(|01\rangle-|10\rangle)$
should be completely stabilized against decoherence. 

{\em{Example 5}}:
${\cal H}= {\cal H}_c^N\otimes {\cal H}_{E},$ and 
$ H=\sum_{i=1}^N H_i,$ where $ H_i$ has non-trivial action only on 
${\cal H}_c^i\otimes {\cal H}_E.$
To make this system ${\cal S}_N$-invariant one only needs
to consider the  subgroup ${\cal Z}_N\subset {\cal S}_N$
of cyclic permutations [acting on ${\cal H}_c^N$].
Indeed if $H_i= \sum_l (X^l_i\otimes B^l_i + \mbox{h.c.})$
then $\tilde H= 
\sum_l ( X^l\otimes B_l +  \mbox{h.c.})$
where $ A^l=: \sum_{i=1}^N A_i^l,\,(A=X,\,B).$

The latter  example show that when $H$ has some symmetry  from the beginning
one can achieve full $\cal G$-invariance by resorting to an ancillary space
{\em  smaller than} ${\bf{C}}{\cal G}.$
Here one just needs an ancilla that is exponentially
smaller than  ${\bf{C}}{\cal S}_N.$
This result can be extented to the case in which $H$ is ${\cal G}^\prime$-invariant
where ${\cal G}^\prime\subset{\cal G}$ is a (normal) subgroup.
To exemplify this situation let us consider a lattice Hamiltonian $H$
over a regular polygon $\cal P$ with $N$ vertices. Suppose  $H$ to be invariant
with respect to the group ${\cal Z}_N$ of cyclic permutations of the  sites of $\cal P$.
To make $H$ invariant under the full group ${\cal D}_N$ of isometries of $\cal P$ 
just a two-dimensional ancilla (one qubit) is required. This stems
from the fact that the coset space 
${\cal G}/{\cal G}^\prime={\cal D}_N/{\cal Z}_N\cong {\cal Z}_2$ has order two.

The symmetrization procedure can be used for getting rid of  unwanted terms
in a system Hamiltonian. Let us suppose that $H = H_0+ H_1$
where $H_0$ is $\cal G$-invariant and $H_1$ transforms according the $i$-th row of  the $J$-th irrep of $\cal G,$
i.e., $\rho_g^\dagger\,H_1\,\rho_g=\sum_j \rho^J_{ji}(g)\, H^j.$
Then from the orthogonality relation $\sum_g \rho^J_{ji}(g)=0$ \cite{CORN} 
one obtains $\pi_{\tilde\rho}(H_1)=0$
and therefore
 $\tilde H=H_0.$
This result can be in principle used for suppressing decoherence 
in a  quantum computer.
This issue is illustrated in the next two examples that deal respectively 
with  $N$ qubits and with an harmonic oscillator
coupled with a dissipating  environment\cite{LLO}, \cite{VITO}.

{\em Example 6} Let ${\cal H}={\bf{C}}^{2\,N}\otimes{\cal H}_E,\, 
H=H_0+ \sum_{i=1}^N  ( \sigma^+_i\otimes E_i+ \sigma^-_i\otimes E^\dagger_i),$
and $\rho\colon {\cal Z}_2\mapsto \{{\openone},\,\sigma^{z\,\otimes N}\otimes {\openone}\}.$
Suppose that $H_0$ is ${\cal Z}_2$-invariant,  
from $\sigma_i^z\,\sigma_i^\alpha\,\sigma_i^z=-\sigma_i^\alpha\,(\alpha=\pm)$ it follows
that $H_1=\sum_i (\sigma^+_i\otimes E_i +\sigma^-_i\otimes E_i^\dagger)$ transforms according the antisymmetric irrep.
The result can be easily generalized to different kind of interactions,
for example if ${\cal G}$ is the Pauli group
$ \{ {\openone},i\sigma^x,\,i\sigma^y,\,i\sigma^z\}$
and $\rho$ the $N$-fold tensor representation [$\rho\colon \sigma_\alpha\mapsto 
\sigma_\alpha^{\otimes\,N}$]
one can eliminate general couplings with the form $\sum_{i\alpha={x,y,z}}\sigma^\alpha_i
\otimes E_i^\alpha.$
In particular if only the $\sigma_i^z$'s are present
one is dealing with a purely decohering environment.
Since $\sigma_\alpha^{\otimes\,N}\sim\otimes \exp(i\,\pi\sigma_\alpha)=
\exp(i\,\pi\sum_{i=1}^N\sigma_i^\alpha),$ 
here the $\rho_g$'s corresponds to collective ``$\pi$-pulses'' along the $\alpha=x,y,z$ directions. 
Notice that the invariance of free Hamiltonian
holds for operators with the form $H_0=H_S\otimes\openone+\openone\otimes H_B,$ 
where $H_S=\sum_{ij} G_{ij}
 {\bf{s}}_i\cdot  {\bf{s}}_j,$ 
that can be used for providing the conditional dynamics required 
for, along with single-qubit operations,  universal  quantum computation.

{\em Example 7} Let ${\cal H}={\cal H}_B\otimes{\cal H}_E$ where
${\cal H}_B=\mbox{span}\{|n\rangle\}_{n=}^\infty
$ is a single boson mode Fock space 
(with field operator $a$) and
${\cal H}_E$ an environment state-space.
We set $H_0= \omega\,a^\dagger\,a\otimes{\openone}+ {\openone}\otimes H_E,\,
H_1=a^\dagger\otimes E+ a\otimes E^\dagger.$ Now the relevant representation is
$\rho\colon{\cal Z}_2\mapsto\{ {\openone},\, \exp( i\,\pi\,a^\dagger\,a\otimes{\openone})\}.$
Once again the system-environment interaction Hamiltonian $H_1$ is averaged away in that
it has odd parity i.e.,
$ \exp( i\,a^\dagger\,a)\,a\, \exp( i\,a^\dagger\,a)=-a.$ 
Notice that this example corresponds to {\em Ex. 3} being the 
 subspace given by the {\em even} sector
${\cal H}_e=:\mbox{span}\{|2\,n\rangle\}_{n=0}^\infty={\cal H}^\rho_{inv}.$ 

The strict relation with the frequent pulse control of  
decoherence proposed in refs. \cite{LLO} and \cite{VITO} should be clear.
In fact this analogy allows to reformulate the whole symmetrization strategy
by  a procedure that {\em does not} resort to any ancilla and measurement.

{\em Unitary symmetrization. }
Let $\rho_i=\rho_{g_i}\,i=1\ldots,|{\cal G}|$ the group representatives.
Consider a time interval $\delta t_N=t\,(N\,|{\cal G}|)^{-1}$ and let $\delta U_N=\exp(-i\,\delta t_N\,H)$
then apply the following sequence of transformations
\begin{eqnarray}
& &U_N(t) = \prod_{i=1}^{|{\cal G}|}\rho_i^\dagger\,\delta U_N\,\rho_i=
\prod_{i=1}^{|{\cal G}|}e^{-i\,\delta t_N\,\rho_i^\dagger\,H\,\rho_i}
 \stackrel{N\mapsto \infty}{\simeq}\nonumber \\
& &
\exp\left(-i\frac{t}{N\,|{\cal G}|}\sum_{i}g_i^\dagger\,H\,g_i\right)  
{\simeq} \exp\left( -i\frac{t}{N}\tilde H\right),
\label{limit}
\end{eqnarray}
implying $U(t)=\lim_{N\mapsto\infty} [U_N(t)]^N= \exp\left(  -i\,t\,\tilde H\right ).$
Notice that here, for simplicity, we assumed that the unitaries $\rho_g$'s
can be realized in a vanishingly small amount of time in which the evolution induced by $H$
is negligible.
A detailed analysis of
the physical requirements needed in order to achieve
the  limit (\ref{limit})
can be found, for specific cases, in refs. \cite{LLO}, \cite{VITO}.

This unitary realization of $\cal G$-symmetrization
could be, from the point of view of feasibility, much  better than the procedure
based on iterated measurements.
Indeed the latter implies extra space resources, the capability of carrying on 
unitary transformations (the $W_\rho$'s) that are possibly highly non-trivial
and iterated measurements. 
 In these respects the first procedure resembles
the  Error Correction techniques \cite{ERR}.

Our analysis  sheds light on the  structure underlying the decoherence-suppression strategies:
the application of the symmetrization  procedure can be viewed as an {\em harmonic}
 filter that selects out the decoherence-inducing  part of the Hamiltonian
in view of its representation-theoretic structure.
This phenomenon is connected to the fact that, in the above examples,
the symmetry content of a subspace is related to the number of ``elementary excitations''
contained in it. Since the interaction Hamiltonian $H_1$ describes the exchange
of such elementary objects, it couples different symmetry sectors,
therefore it cannot belong to the set of $\cal G$-invariant operators.

The experimental realization of the scheme analysed in this letter
is in general extremely demanding.
One should able to perform unitary operations (and  measurements), each one requiring a
time  $\tau,$ with a frequency $\nu$ 
much greater than the one associated to the fastest time scale of the  evolution generated by $H.$
For instance in case {\em 1} one must have $\tau^{-1}\gg \nu\gg \omega_c,$ where $\omega_c$
is the bath frequency cut-off (see refs. \cite{LLO}, \cite{VITO}).
In the scheme involving measurements one could  turn on, for a time $\tau$ and with frequency $\nu,$
 the Hamiltonians $H^\rho(t)= f(t)\,K^\rho,$
where $\int_0^\tau dt\, f(t)=1.$ 
If these requirements are not exactly fulfilled  one obtain a partial symmetrization
for which, as far as the last examples are concerned, the  noise is just reduced rather
than eliminated.
Moreover, for general ${\cal G}$ and $\rho$ 
the ``pulses'' $\rho_g$'s will be quite difficult to implement.
Roughly speaking,
this amounts to the capability of switching on the Hamiltonians $h_\rho^g$
that in general will correspond to  non-trivial collective interactions.
On the other hand all the up-to-date  proposals  for maintaining coherence in a quantum computer
are known to be quite challenging from the point of view of implementation.
Conceptually it is intriguing to realize that all these techniques 
have at their root a group-theoretic structure. 

I thank 
M. Rasetti for stimulating discussions and critical reading of the manuscript,
Elsag, a Finmeccanica Company, for financial support.
%END 
%%%%%%%%%%%%%%%%%%%%%%%%%%%%%%%%%%%%%%%%%%%%%%%%%%%%%%%%%%%%%%%

\end{multicols}
\end{document}